\documentclass[twocolumn,preprintnumbers,amsmath,amssymb]{revtex4}
\usepackage{graphicx}
\usepackage{subfig}
\usepackage[format=plain,justification=raggedright,singlelinecheck=false,font=small,labelfont=bf,labelsep=space]{caption}

\begin{document}

\title{Rearrangements during slow compression of a jammed two-dimensional emulsion}
\author{Xin Du}
\affiliation{Department of Physics and Astronomy, Widener University, Chester, PA, 19013}
\email{xdu@widener.edu}

\author{Eric R. Weeks}
\affiliation{Department of Physics, Emory University,
Atlanta, GA 30322, USA}
\email{erweeks@emory.edu}

\date{\today}

\begin{abstract}

As amorphous materials get jammed, both geometric and dynamic heterogeneity are observed. We investigate the correlation between the local geometric heterogeneity and local rearrangements in a slowly compressed bidisperse quasi-two-dimensional emulsion system.  The compression is driven by evaporation of the continuous phase, and causes the area packing fraction to increase from 0.88 to 0.99.  We quantify the structural heterogeneity of the system using the radical Voronoi tessellation following the method of [Rieser {\it et al.}, {\it Phys. Rev. Lett.} {\bf 116}, 088001 (2016)].  We define two structural quantities characterizing local structure, the first of which considers nearest neighbors and the second of which includes information from second nearest neighbors.  We find that droplets in heterogeneous local regions are more likely to have local rearrangements.  These rearrangements are generally T1 events where two droplets converge toward a void, and two droplets move away from the void to make room for the converging droplets.  Thus the presence of the voids tends to orient the T1 events.  The presence of a correlation between the structural quantities and the rearrangement dynamics remains qualitatively unchanged over the entire range of packing fractions observed.

\end{abstract}

\maketitle


\section{Introduction}

There are a variety of soft amorphous solids:  for example, emulsions, foams, and colloids.  An emulsion is composed of droplets of one liquid immersed in a second immiscible liquid, with the droplets coated with surfactant molecules to prevent droplet coalescence.  A foam is similar except with gas bubbles.  A colloid is composed of solid particles in a liquid.  In all of these cases, these materials become ``jammed'' as the packing fraction of the systems increases \cite{liu98}.  The control parameter is the volume fraction $\phi$ (or area fraction for two-dimensional systems).  For a foam, for example, if the gas volume fraction is above about $\phi_J \approx 0.65$, the foam has a yield stress and can form a pile on the table \cite{liu10,vanhecke10,kraynik88}; this identifies $\phi_J$ as the jamming volume fraction.  In contrast to solids which crystallize, the structure of these jammed solids is amorphous and thus spatially heterogeneous \cite{vanblaaderen95,royall08}.

Even though such materials have a yield stress, emulsions and foams are composed of soft components (bubbles, droplets) and thus can be forced to flow at high $\phi > \phi_J$ \cite{mason95emul,dollet07,dollet10,jones11,chen12,dollet15jfm,vasisht18,cao21}.  Given the structural disorder and the fact that flowing requires the component particles to move past one another, it is not surprising that the flow is highly disordered and involves particles moving and rearranging collectively \cite{chen10}. It also is sensible that there can be some connection between local structure and local dynamics \cite{royall08, tsamados09, manning11,jack14, cubuk15}. Much of the prior work was done at constant density, leaving an open question as to what are useful ways to characterize structure that are density-independent (although see Ref.~\cite{tah22} for recent work applying machine learning across a range of densities).  

\begin{figure}[b]
\centering
    \subfloat{{\includegraphics[width=3.9cm]{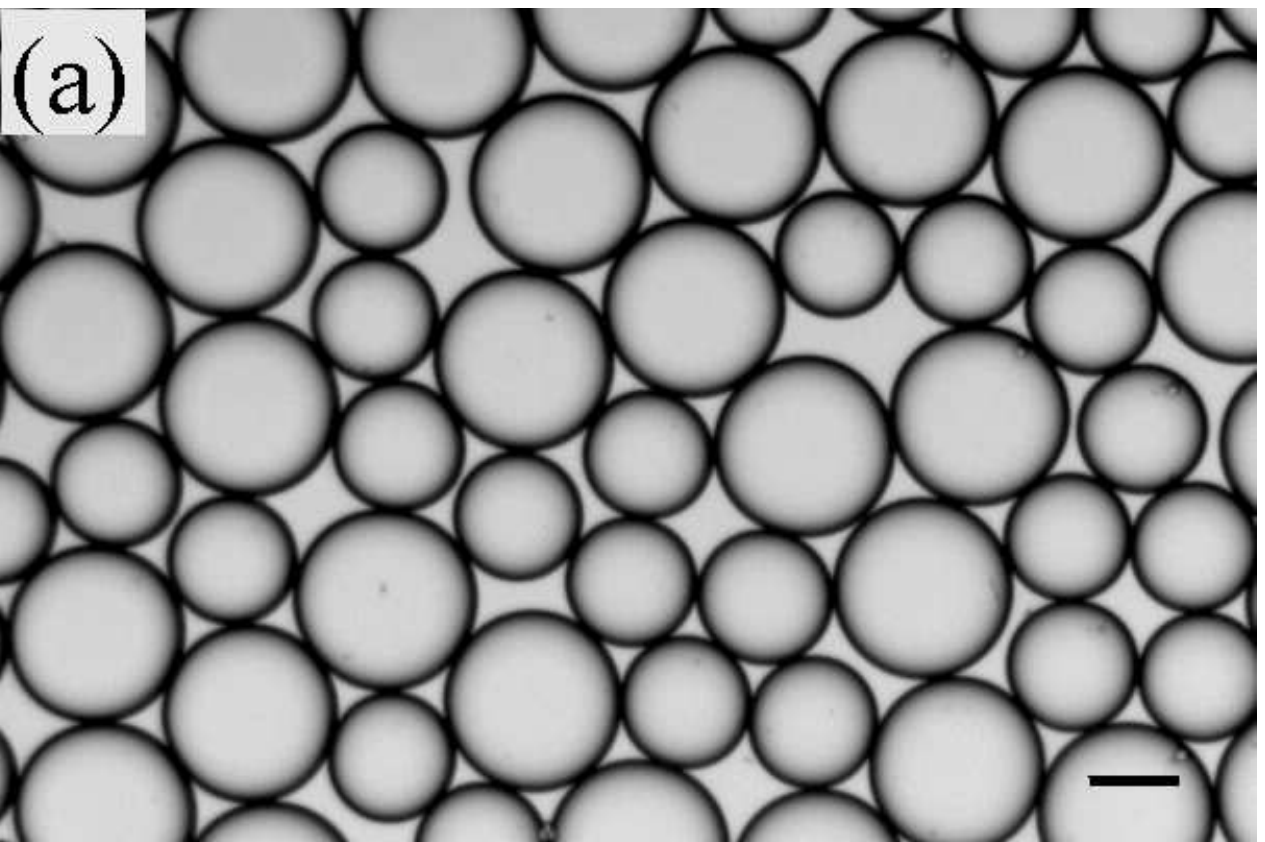} }} 
    \subfloat{{\includegraphics[width=3.9cm]{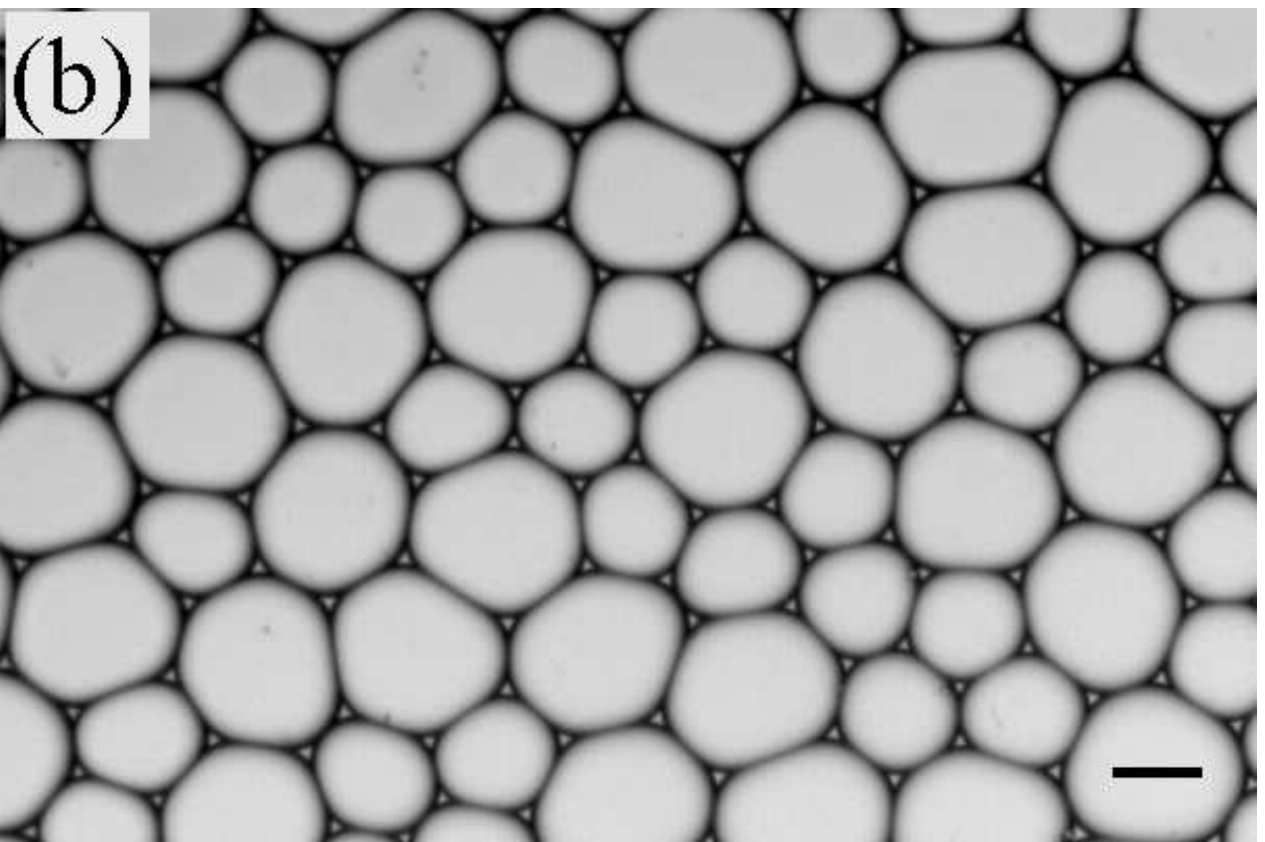} }} \ 
\caption{ The images of a portion of the emulsion sample at $t = 0$~min with $\phi =
0.88$ (a), and at $t = 100$~min with $\phi \rightarrow 1$ (b). The field of view in (a) and (b) is 1.76 $\times$ 1.44 mm$^2$. The scale bar represents 200~$\mu$m.}
\label{fig:images}
\end{figure}
In this paper, we study a quasi-two-dimensional emulsion sample composed of oil droplets in water.  The water is allowed to slowly evaporate so that we can study the rearrangements that occur as the area fraction increases from $\phi = 0.88$ (just above jamming) to $\phi = 0.99$ (overjammed), as shown in Fig.~\ref{fig:images}.  We characterize the structural heterogeneity of the jamming emulsions using structural quantities proposed by Rieser {\it et al.} \cite{rieser16}.  These quantities derive from the radical Voronoi tessellation, and account for the structure of first nearest neighbors, or both first and second nearest neighbors.  We observe correlations between the rearrangements droplets exhibit, and structural measures indicating voids between droplets or other structural inhomogeneities.  As is expected, droplets are more likely to move if they are near voids.  However, in particular we note that droplets tend to move with definite orientations relative to those voids.  Intriguingly, we demonstrate that the correlations between structure and motion stay qualitatively (and in some cases quantitatively) similar even as the area fraction changes significantly.  This shows the utility of the structural measures introduced by Rieser {\it et al.}

\section{Experimental methods}

We use emulsions for several reasons.  First, they are relatively inexpensive samples to produce \cite{bibette99,derkach09}.  Second, we can achieve values of $\phi$ well above jamming as the droplets can deform \cite{desmond13}.  Third, we can exploit evaporation to slowly change $\phi$ {\it in situ}.  Fourth, unlike foams, the emulsion samples do not coarsen -- the volume in each oil droplet remains constant throughout the experiment.

\begin{figure}
\centering \includegraphics[width=0.6\columnwidth]{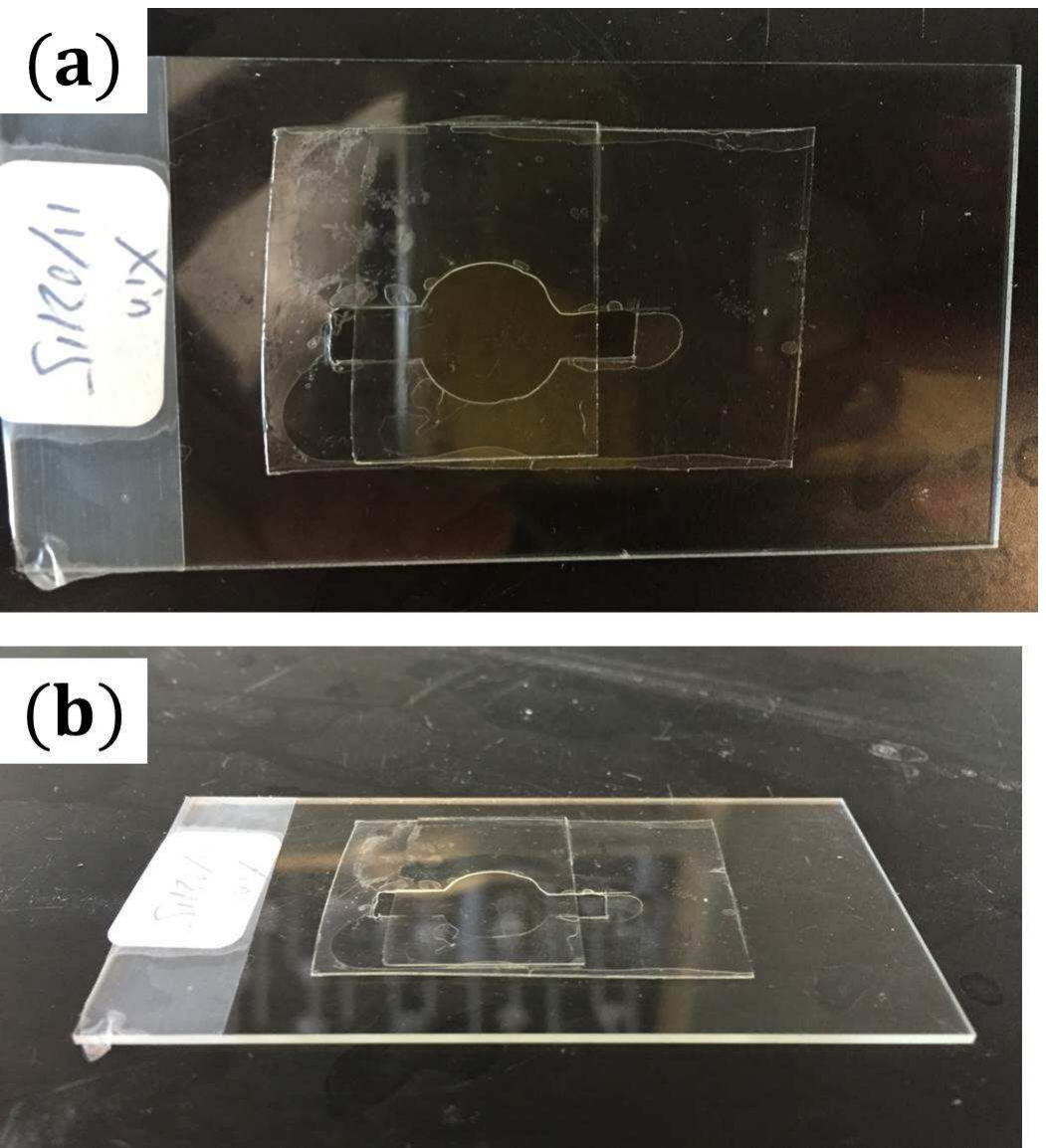}
\caption{\label{chamber} (a)
Top view and (b) side view of the sample chamber. Water slowly
evaporates from the two chamber openings at left and right.  The slide is 75~mm long for scale.} \end{figure}

Our emulsions are mixtures of water and silicon oil droplets using Fairy detergent (mass fraction 0.025) as a surfactant to stabilize the droplets against coalescence. The oil droplets are generated using a standard co-flow microfluidic technique \cite{shah08}. We make bidisperse emulsions with two batches of monodisperse droplets at a volume ratio of about 1:1; thus the small droplets are more numerous.  The droplets are then placed in a quasi-two-dimensional (quasi-2D) sample chamber and deform into pancake shapes \cite{desmond13,chen12}.  We make our sample chamber with two 25 mm $\times$ 75 mm microscope glass slides separated by 100~$\mu$m thick spacers (transparency film) sealed with epoxy. The spacers with the epoxy create a gap of 120~$\mu$m between the two glass slides.  During construction of the sample chamber, we use clamps to press the two slides together to try to squeeze the epoxy to a uniform thin layer.  The final chamber thickness varies slightly across the sample chamber, but by no more than $\pm 3$~$\mu$m.  We do not observe any droplet motions that appear connected to the thickness variability.  The spacers are cut with a central circular space to contain the sample, and two smaller openings at the sides for evaporation; see Fig.~\ref{chamber}. 

The mean 2D diameters of the small droplets and the large droplets are 265~$\mu$m and 379~$\mu$m respectively, resulting in a size ratio of about $1:1.43$.  The individual droplet species have a polydispersity (standard deviation of droplet radii divided by mean radius) of 4\% and 2\% for small and large droplets.  The diameters are well defined at the start of the experiment ($\phi=0.88$), and as the droplets compress into polygons, the diameter defined as $d=\sqrt{4A/\pi}$ stays similar for each droplet as compared to its earlier value.  To initialize the experiment, the two batches of droplets are premixed in a vial, and then added to the sample chamber via a pipette.  In practice this pre-mixing results in reasonably well-mixed samples as observed in the sample chamber.  The data to be presented are from one experimental run.  In this sample, there is additionally one unusually large droplet with diameter 883~$\mu$m, formed likely due to coalescence of droplets prior to adding the sample to the microscope slide.  All droplet diameters are larger than the sample chamber thickness, ensuring that the sample is quasi-2D.  (Several other experiments were conducted with generally similar results to what is reported below, but had unacceptable net flows driven by the evaporation process, and so were discarded from our analysis.)

After the sample chamber is filled, it is placed on a microscope for imaging with a $5\times$ objective lens.  A 1280 $\times$ 1024 pixel image is recorded every 30 seconds with a camera mounted on the microscope for 100 minutes.  The field of view is 6.07 $\times$ 4.85 ${\textup{mm}}^2$ and is comprised of over 300 droplets.   We track the trajectories of all the droplets using standard techniques \cite{crocker96,chen12}.  Our particle position uncertainty is $\pm 2.4$~$\mu$m.  Given that we need to see the entire droplet outline to successfully analyze its center of mass and shape, we do not track droplets which overlap the edges of the images.  The number of droplets tracked is $\sim 170$ at the start of the experiment and as the sample evaporates and droplets are pressed together, the number rises to $\sim 230$ by the end of the experiment.

To determine the area fraction, we do direct image analysis of the raw images.  First, we fill in all droplet areas that have been previously identified.  Second, we additionally fill in all pixels darker than the threshold used for the original droplet identification:  this thus fills in all pixels except those which are in voids.  The area fraction is then the area of the filled pixels divided by the area of the radical Voronoi polygons enclosing the identified droplets; the radical Voronoi polygon method will be described below.  Overall, given the optical distortion of the true position of the droplet boundaries, there is some uncertainty of the droplet area leading to uncertainty of $\phi$ \cite{poon12}.  We therefore estimate a likely systematic uncertainty on $\phi$ of $\pm 0.02$.  Note also that this treats the droplets as strictly two-dimensional; we do not account for the curvature at the droplet edges due to the confining glass plates.  At the high area fractions we consider such that droplets are pressed together, the details in the direction of the optical axis -- that is, perpendicular to the 2D image -- are unlikely to matter \cite{desmond13}.

\begin{figure}
\centering
    \subfloat{{\includegraphics[width=8cm]{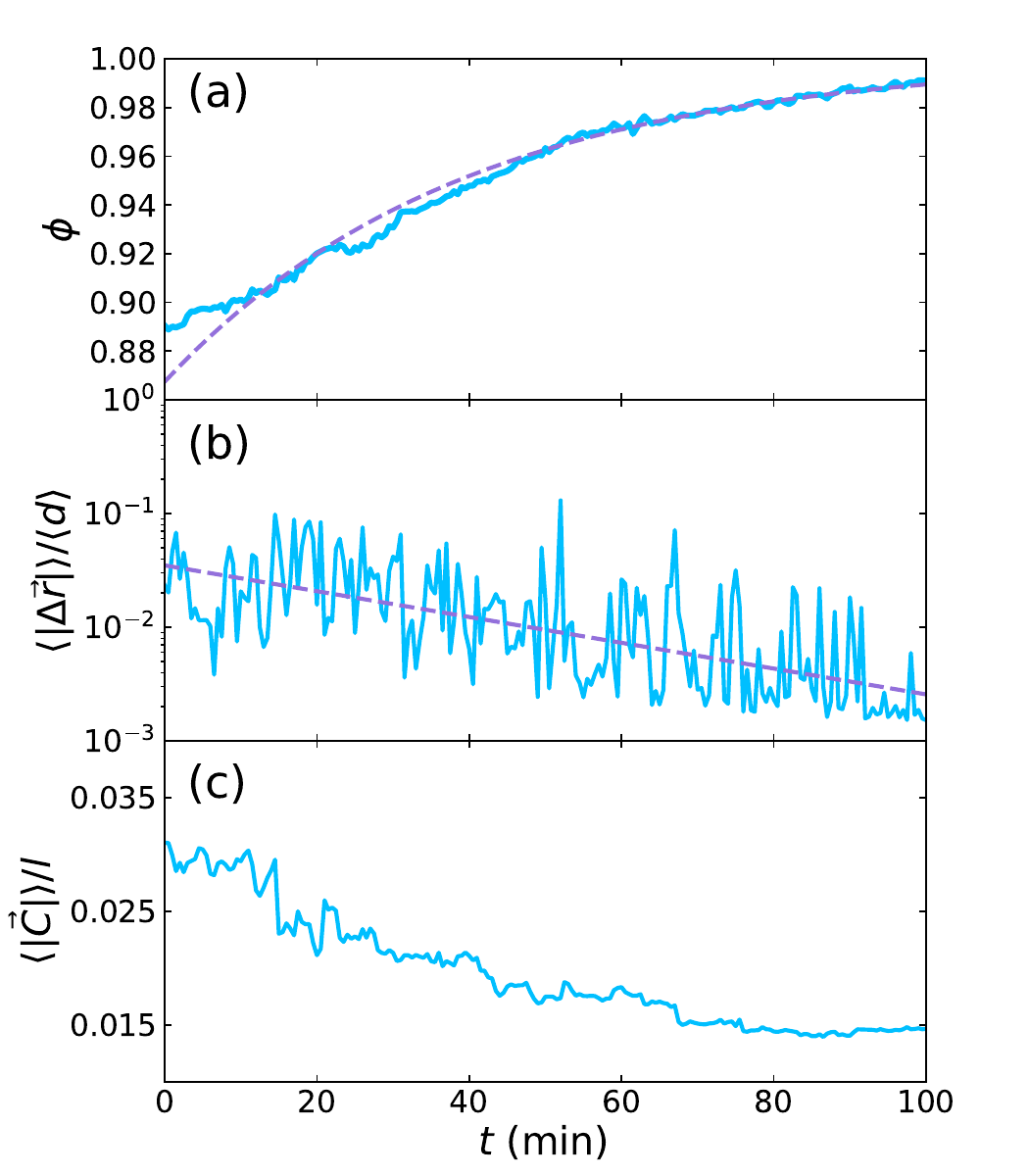} }}
\caption{Various experimental quantities change over time.   (a) The measured packing fraction $\phi$ as a function of time. The dashed line fit is the function $\phi(t) = 1 - (1-\phi_0)\exp(-t/\tau)$ with $\phi_0=0.867$ and $\tau = 39$~min.(b) The average cage-relative displacement of the droplets in each pack, rescaled by the mean diameter of all the droplets $\langle d \rangle = 290$~$\mu$m, as a function of time with a logarithmic vertical axis. For this exponential fit, $A = 0.035$ and $\tau = 38$~min. (c) The average length of the Voronoi anisotropy vector $\langle |\vec{C}| \rangle$, normalized by the interparticle spacing $l$., as a function of time.  }
\label{fig:phivstime}
\end{figure}

To slowly drive rearrangements, we allow water to evaporate from the two sample chamber openings.  Over the course of the experiment, the sample transitions from less jammed to well overjammed with the packing fraction $\phi$ increasing from 0.88 to 0.99.  (The sample is expected to jam at $\phi \approx 0.84$ \cite{bolton90,durian95,koeze16}.)  Images from the beginning and end of the experiment are shown in Fig.~\ref{fig:images}(a,b).  In panel (a), the droplets are random close packed and barely deformed at $t = 0$~s. At $t = 100$~min in Fig.~\ref{fig:images}(b), the droplets are deformed into non-circular shapes, some of which are close to polygons.  As the area fraction slowly increases, droplets deform and occasionally rearrange.  The area fraction as a function of time is plotted in Fig.~\ref{fig:phivstime}(a).  The data for $\phi(t)$ are well fit by an exponential function, although we do not know of a particular reason that it should be an exponential.  It is plausible that as $\phi$ increases, the remaining small amount of water finds it harder to travel through the smaller channels between droplets, thus decreasing the evaporation rate.  The fitted decay time of 39~min quantifies the slow speed at which the sample changes.  This time scale can be compared to previously observed droplet motion time scales:  droplet rearrangements in a slowly driven quasi-2D emulsion similar to our sample occurred on time scales of a few seconds \cite{chen12}.  This suggests that the area fraction of our experiment increases quite slowly compared to the time scale for any rearrangements that are caused by that area fraction change.

\section{Results}

Our goal is to understand how local structure influences where the occasional droplet rearrangements occur.  We first need to quantify rearrangements and do so by calculating the displacement of each droplet, using the time scale $\Delta t=30$~s.  There is a difficulty with this definition: occasionally large groups of droplets all move collectively -- but these are motions relative to the camera, not relative to each other. Accordingly, we quantify the motion using the concept of the cage-relative displacement \cite{mazoyer09}.  The cage-relative displacement is defined as the displacement of a droplet relative to the average displacement of its nearest neighbors, where we define nearest neighbors through a radical Voronoi tessellation (discussed further below).  Thus, these displacements quantify rearrangements of droplets relative to their neighbors rather than collective motions. In the remainder of this paper, we will use $\Delta \vec{r}$ to refer to the cage-relative displacements.  Figure~\ref{fig:phivstime}(b)
is the plot of the average cage-relative displacement of the droplets rescaled by the mean diameter of all the droplets $\langle d \rangle = 290 \mu$m. In general the average local movement of the droplets, $\langle |\Delta \vec{r}|\rangle/\langle d \rangle$, gradually slows down, but with many fluctuations. In some moments the movements are dramatically larger than the others, revealing the temporal dynamic heterogeneity through the entire compression process.


\begin{figure}
\centering
\includegraphics[width=0.95\columnwidth]{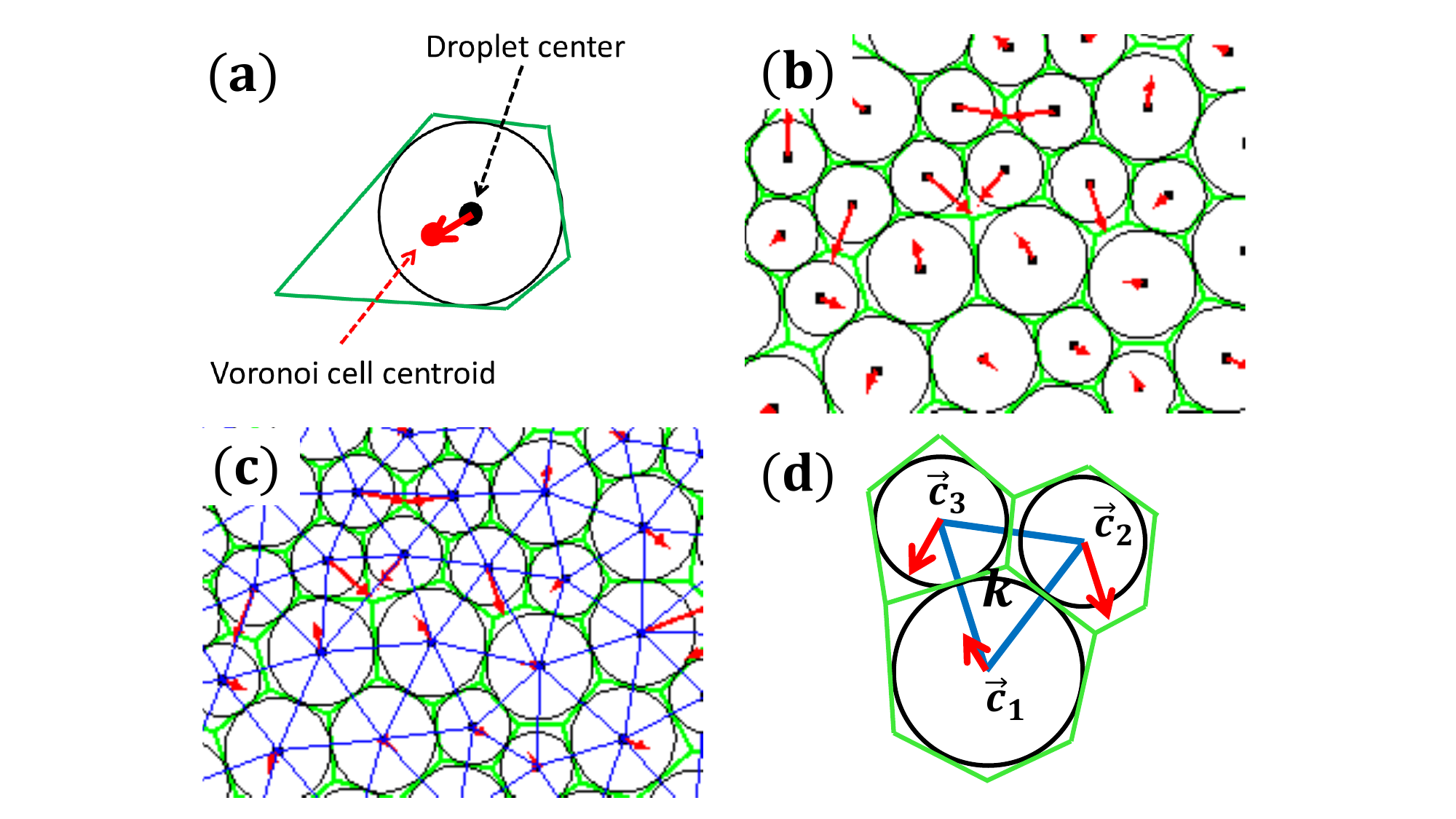}
\caption {\label{voronoicell}
(a) The polygon is the radical Voronoi cell for the droplet. The black dot is the droplet center and the offset red dot is the radical Voronoi cell center. Typically, these two centers are not at the same position. The anisotropy vector is shown as the red arrow which points from the particle center to the Voronoi cell center. (b) The anisotropy vector field is demonstrated by the arrows for each
droplet. (c) The anisotropy vector field (arrows) and radical Delaunay triangulation. (d) An example of one radical Delaunay triangulation with three corresponding droplets and the anisotropy vectors. In (b), (c), and (d), to make the arrows visible, we magnify the length of the vectors by a factor of 30.
}
\end{figure}

We next wish to quantify the local structure so as to determine what features of the structure influence the cage-relative motions.  We quantitatively characterize the local geometric heterogeneity of the system based on radical Voronoi tessellation.  The radical Voronoi tessellation is a standard partition of space which takes into account the radius of each droplet, and which highlights the closest free space near each droplet. {Specifically, this geometric analysis considers each droplet as a circle with diameter $d = \sqrt{4 A / \pi}$ based on its area, and centered on the center of mass of the droplet.  Each droplet is surrounded by a polygon comprised of all points closer to that circle's boundary than to any other circle's boundary.  The edges of these polygons correspond to points for which the length of tangents to each adjacent circle are equal.  These are the radical Voronoi polygons which tile space, and are} displayed as the polygons in Fig.~\ref{voronoicell}. Then, we define the anisotropy vector as the vector pointing from the droplet centroid to the Voronoi cell centroid, displayed as the arrows in Fig.~\ref{voronoicell}. The magnitude of anisotropy vector is zero if the free space is homogeneously distributed near a droplet. Thus, the length of anisotropy vector, $|\vec{C}|$, represents the geometric heterogeneity of the nearest neighborhood of a droplet.

As the sample area fraction increases over time, the average vector lengths decrease.  However, there are two potential reasons for this decrease:  the sample could be restructuring in interesting ways, or the droplets could merely be moving closer together (consistent with the increasing area fraction) and thus the size of the Voronoi cells decreases.  To remove the trivial geometric effect of droplets getting close together, we divide $|\vec{C}|$ by the typical distance between droplets $l$, which is the square root of the inverted number density of the droplets in each packing.  Thus, $|\vec{C}|/l$ quantifies the local geometric heterogeneity of a droplet. In our experiment $|\vec{C}|/l$ of the droplets ranges from $0$ to $0.09$.  The average of this quantity decreases  with time, as shown in Fig.~\ref{fig:phivstime}(c), although the functional form of this decrease is unclear. Given the normalization by $l$, this remaining time dependence reflects that the structure becomes more homogeneous as $\phi$ increases:  droplets are more likely to reside close to the centers of their Voronoi polygons.

\begin{figure*}
\centering
\includegraphics[width=1.7\columnwidth]{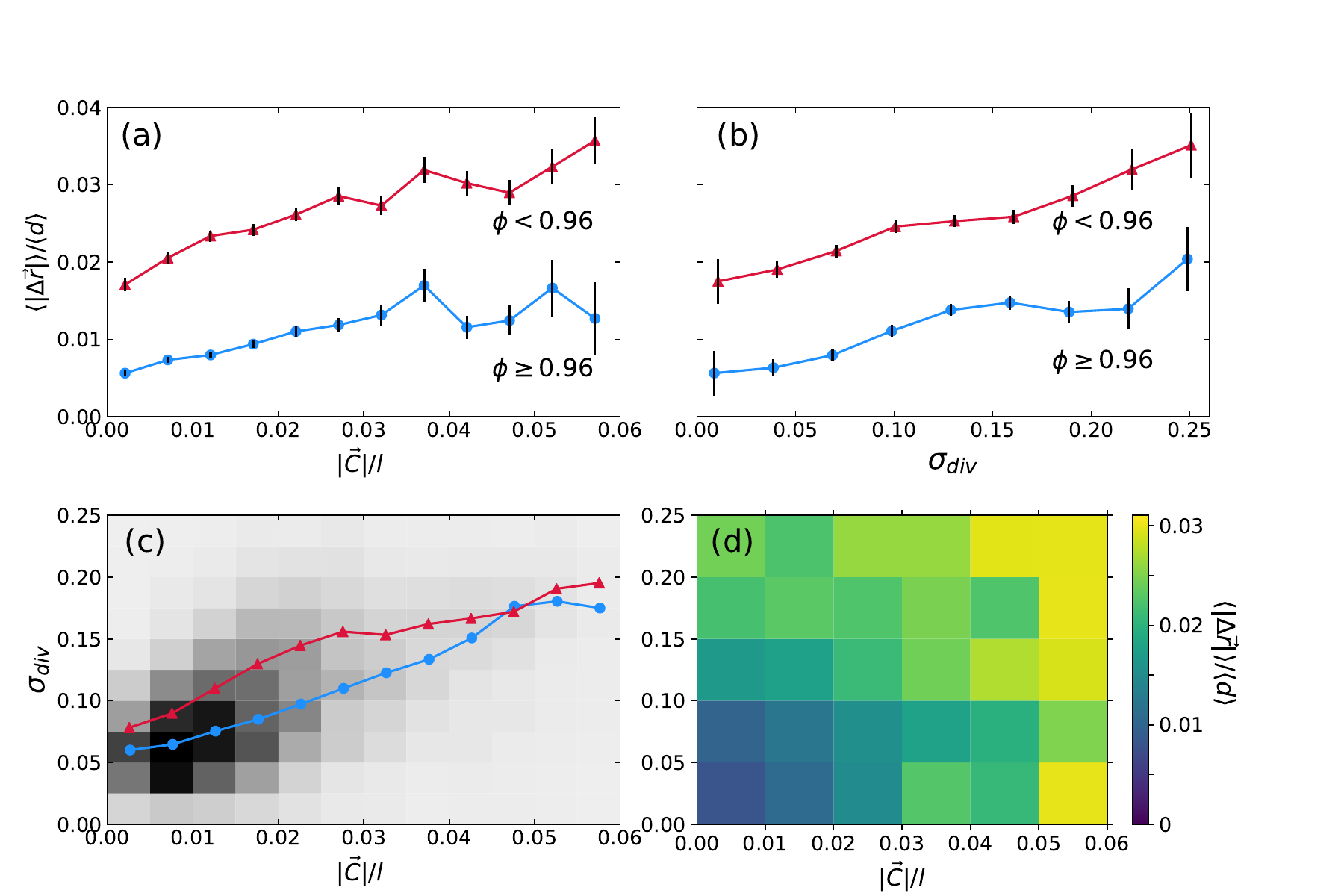}
\caption{\label{drcage} The relationship between the droplets' local movements and the features of their local structure in both less jammed packing with $\phi<0.96$ (red triangles) and overjammed packings with $\phi \geq 0.96$ (blue circles). (a) Plot of the mean cage-relative displacements as a function of the normalized Voronoi vector length. Each data point corresponds to data averaged over at least 100 droplets with $|\vec{C}|/l$ falling in a certain range. The black error bars are the uncertainties of the mean (the standard deviation of the $N$ values going into each average, divided by $\sqrt{N}$). (b) Plot of the mean cage-relative displacements as a function of the standard deviation of the local divergence of the anisotropy vector field.  Each data point is an average including at least 100 droplets with $\sigma_{\rm div}$ falling in a certain range. The black error bars are the uncertainties of the mean. (c) Plot of $\sigma_{\rm div}$ as a function of $|\vec{C}|/l$.  The gray-scale heat map indicates the counts (over the full duration of the experiment) seen at each location, ranging from a low of $4-20$ counts in the upper-left/lower-right corners, to a high of $\sim 2400$ for the darkest bins.  The red triangles ($\phi < 0.96$) and blue circles ($\phi \geq 0.96$) are data points showing the average $\sigma_{\rm div}$ of the droplets with $|\vec{C}|/l$ in a certain range. (d) The heat map of $\langle |\Delta \vec{r}|\rangle/\langle d \rangle$  demonstrating its dependence on $\sigma_{\rm div}$ (y-axis) and $|\vec{C}|/l$ (x-axis).  The lighter shade (yellow online) represents large cage-relative motion, and the darker shade (blue online) represents smaller cage-relative motion, as indicated by the color bar to the right.  The grid size for (d) is coarser than in (c) to ensure that each bin contains at least 60 observations.
}
\end{figure*}

Given that a droplet with larger than average $|\vec{C}|/l$ is in a more geometrically heterogeneous environment -- and in particular has a larger void in the direction of $\vec{C}$ -- we conjecture that the value of $|\vec{C}|/l$ is positively correlated with motion.  Figure~\ref{drcage}(a) shows that this conjecture is true:  droplets with larger $|\vec{C}|/l$ exhibit faster motion on average in both less jammed (red curve with $\phi<0.96$) and overjammed (blue curve with $\phi \geq 0.96$) packings. The droplets' motion, $|\Delta \vec{r}|/ \langle d \rangle$, is the cage-relative displacement of a droplet within $30$~s rescaled by the average diameter of the droplets. During jamming, the droplets move faster if they reside in a heterogeneous nearest neighborhood.  The top curve in Fig.~\ref{drcage}(a) corresponds to the low area fraction data ($\phi < 0.96$, roughly the first half of the experiment) when motion is faster than the high area fraction data shown in the bottom curve.  While the overall magnitude of cage-relative motion decreases over the course of the experiment, the qualitative relation between the magnitude of the motion and the structural quantity $|\vec{C}|/l$ stays similar.  This result is qualitatively similar to prior observations of dense amorphous systems which found connections between locally heterogeneous environments and enhanced likelihood of motion \cite{weeks02,cianci06ssc,matharoo06,kawasaki07}.  Fitting all the data as a linear function of both $\phi$ and $|\vec{C}|/l$ gives us:
\begin{equation}
    |\Delta \vec{r}|/ \langle d \rangle =
    -2.18 (\phi-\phi_J) + 0.345 |\vec{C}|/l + 0.422,
\end{equation}
using $\phi_J = 0.84$.

Rieser {\it et al.} note that the anisotropy vectors ``tend to point in toward locally less well-packed and away from locally more well-packed regions of the packing, reminiscent of sinks and sources in a vector field'' \cite{rieser16}.  Accordingly, they examined the divergence of the anisotropy vector field as a way to characterize voids in the packing, which we will now do with our droplet data.  As is illustrated in Fig.~\ref{voronoicell}(c), the 2D packing is partitioned into many local regions based on the {radical} Delaunay triangulation.  The vertices of a Delaunay triangle are the centers of three neighboring droplets corresponding to three anisotropy vectors, shown in Fig.~\ref{voronoicell}(d). We choose the Delaunay triangle as the local region, and calculate the divergence of the three anisotropy vectors in each Delaunay triangle.  We weight the result with the area $A_k$ of each triangle divided by the mean area $\langle A \rangle$ of all of the triangles, resulting in the final quantity considered by Rieser {\it et al.},
\begin{equation}
Q_{k} \equiv (\vec{\nabla} \cdot \vec{c}) \frac{A_{k}}{\langle A \rangle},
\label{divergence}
\end{equation}
where $\mathbf{c}$ is the field of anisotropy vectors \cite{rieser16}. $Q_{k}$ is a dimensionless quantity that represents the geometric property of the corresponding local region defined by the {radical} Delaunay triangle $k$. Positive (negative) values indicates overpacked (underpacked) regions.  As was true for Rieser {\it et al.}, the distribution of $Q_k$ has a zero mean, is nearly Gaussian, and has a slight negative skewness:  the underpacked areas ($Q_k < 0$) are more probable than would be expected for a perfect Gaussian \cite{rieser16}.  Also in agreement with the results of Rieser {\it et al.}, the standard deviation of the distribution $P(Q_k)$ decreases as the area fraction increases (data not shown).

As displayed in Fig.~\ref{voronoicell}(c), each droplet is surrounded by several {radical} Delaunay triangles, which corresponds to several local divergences $Q_{k}$.  We next compute the standard deviation $\sigma_{\rm div}$ of the several $Q_{k}$ surrounding each droplet.  Each anisotropy vector $\vec{C}$ represents structural information for a droplet and all of its $1^{\rm{st}}$ nearest neighbors, and each $Q_k$ uses information from a droplet's $\vec{C}$ and the $\vec{C}$ of its nearest neighbors.  Accordingly, $\sigma_{\rm div}$ is determined by both the $1^{\textup{st}}$ and $2^{\textup{nd}}$ order nearest neighbors, and represents the structural character of a larger neighborhood than $|\vec{C}|/l$. Large $\sigma_{\rm div}$ indicates that the droplet resides in a heterogeneous local structure and small $\sigma_{\rm div}$ indicates a homogeneous local structure. 

As we can see from Fig.~\ref{drcage}(b), the droplets with larger $\sigma_{\rm div}$ move faster on average in both less jammed and overjammed packs.  The data can be fit to the linear function 
\begin{equation}
        |\Delta \vec{r}|/ \langle d \rangle =
    -2.16 (\phi-\phi_J) + 0.0740 \sigma_{\rm div} + 0.419,
\end{equation}
using $\phi_J=0.84$.  A three variable fit gives
\begin{equation}
    |\Delta \vec{r}|/ \langle d \rangle =
    -2.17 (\phi-\phi_J) + 0.301 |\vec{C}|/l  + 0.0200 \sigma_{\rm div} + 0.420.
    \label{threevarfit}
\end{equation}
The dependence on $\sigma_{\rm div}$ is reasonable:  large $\sigma_{\rm div}$ is another way to note that a particle is in a spatially heterogeneous environment, much as $|\vec{C}|/l$ does.  Thus it is reasonable that the trend of the curves in Fig.~\ref{drcage}(a,b) are similar:  both $|\vec{C}|/l$ and $\sigma_{\rm div}$ influence particle rearrangements.  Equation \ref{threevarfit} shows that the two structural quantities capture slightly different aspects of the motion as both terms have nonzero coefficients.  Given that $|\vec{C}|/l$ changes by $\sim 0.06$ and $\sigma_{\rm div}$ changes by $\sim 0.25$, in Eq.~\ref{threevarfit} the magnitude of the $|\vec{C}|/l$ term is $0.301 \times 0.06 \approx 0.018$ and the magnitude of the $\sigma_{\rm div}$ term is $0.02 \times 0.25 \approx 0.005$.  This suggests the former is a more significant influence on the dynamics than the latter.

Clearly both $\vec{C}$ and $\sigma_{\rm div}$ have a similar influence on particle motion, as seen in Fig.~\ref{drcage}(a,b) and as quantified by Eq.~\ref{threevarfit}.  Both quantities relate to the void structure around a droplet: $\vec{C}$ has geometric information from the nearest neighbors, and $\sigma_{\rm div}$ includes information also from second nearest neighbors. To examine correlations between these two quantities, Fig.~\ref{drcage}(c) shows a a heat map of the two quantities for all droplets at all times; they are indeed correlated.  The lines with red triangles and blue circles show the mean $\sigma_{\rm div}$ as a function of $|\vec{C}|/l$ for lower and higher volume fractions, respectively, and in both cases there is a positive correlation between the two structural quantities.  Are these two quantities redundant for predicting motion, or is one more useful than the other?  To check this, we make a heat map of the magnitude of the cage-relative displacement as a function of both variables in Fig.~\ref{drcage}(d).  The darker data in the lower left corner shows that the smallest magnitude motion is for particles that have both small $|\vec{C}|/l$ and $\sigma_{\rm div}$.  The right portion of the data suggests that $|\vec{C}|/l$ is more influential over droplet motion, in that droplets with small $\sigma_{\rm div}$ but large $|\vec{C}|/l$ are among the most mobile particles. That being said, droplets with small $|\vec{C}|/l$ but larger $\sigma_{\rm div}$ are slightly more mobile than those droplets with both quantities small.

\begin{figure}
\centering
    \subfloat{{\includegraphics[width=0.70\columnwidth]{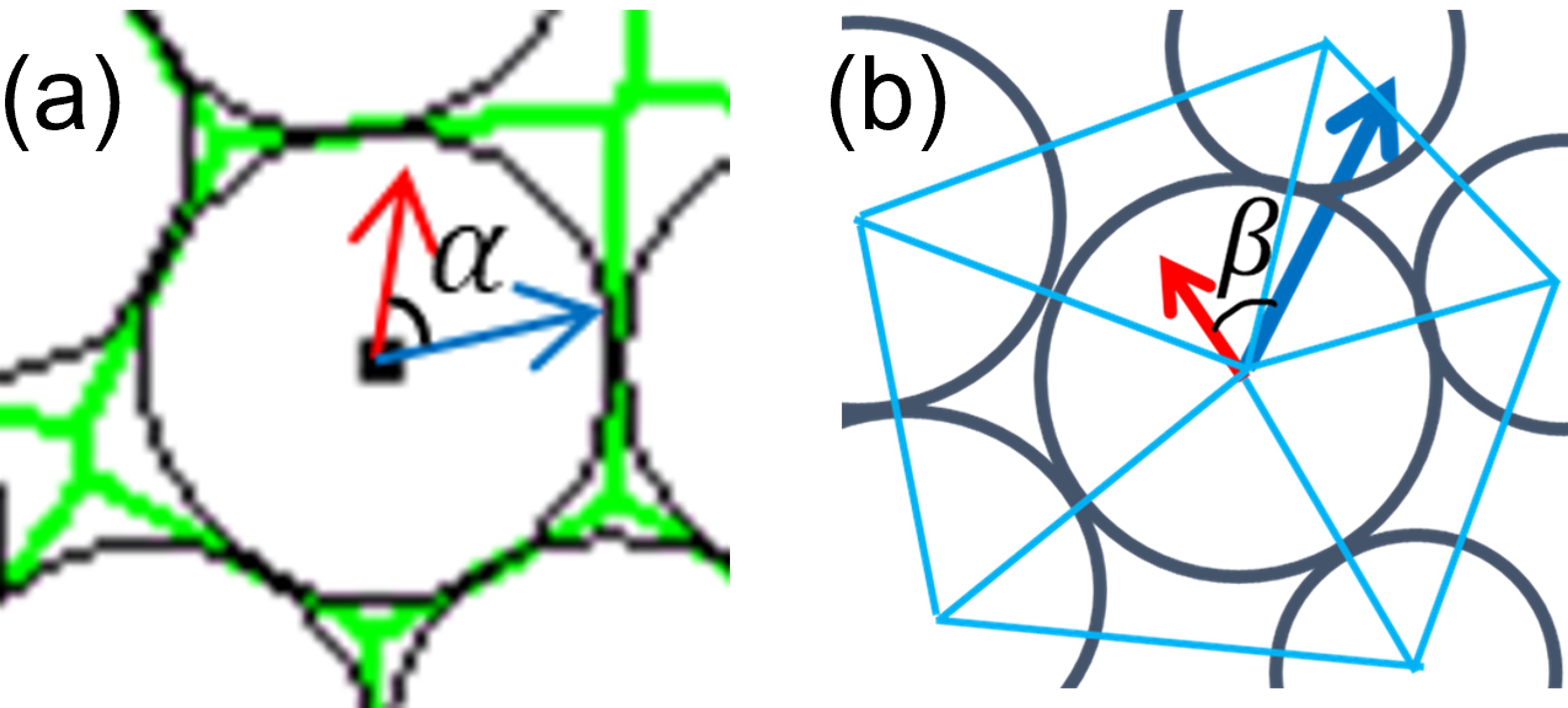} }} \ 
    \subfloat{{\includegraphics[width=0.99\columnwidth]{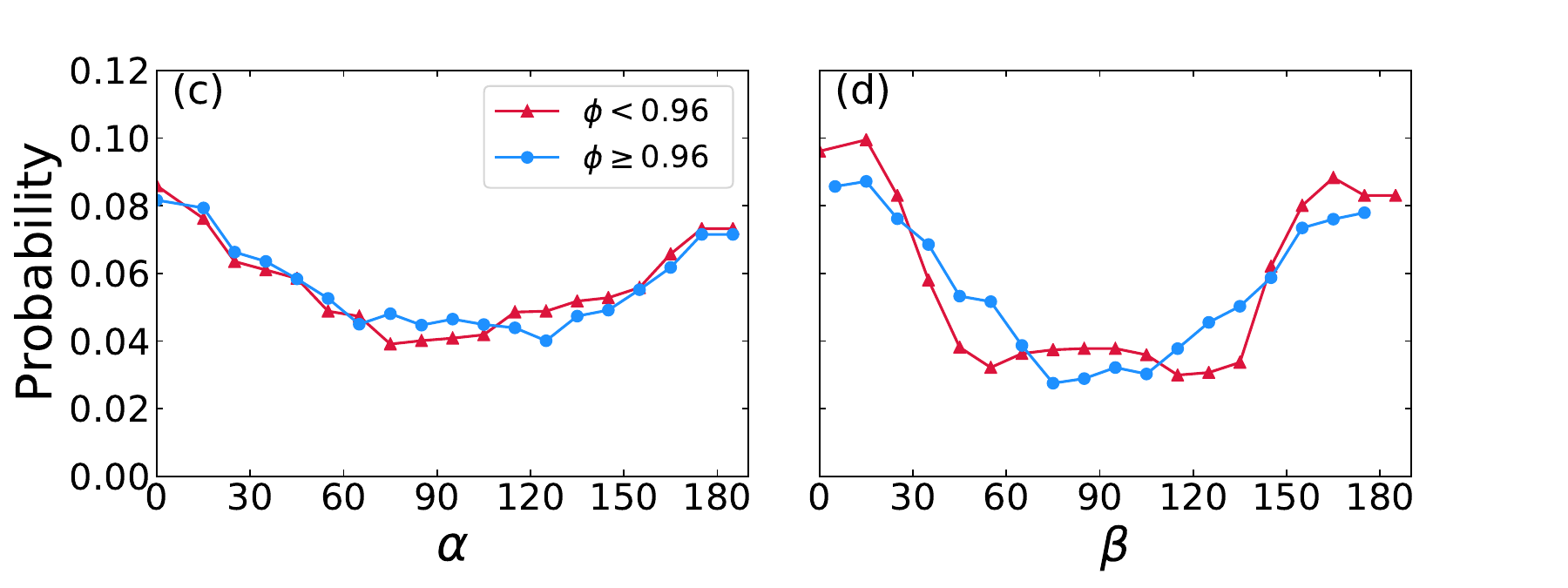}}}
\caption{\label{direction} 
(a) Sketch showing the angle $\alpha$ between the droplet cage-relative displacement direction $\Delta \vec{r}$ (pointing to the right, in blue)  and the Voronoi anisotropy vector $\vec{C}$ (pointing up, in red).  (b) Sketch showing the angle $\beta$ between $\Delta \vec{r}$ (pointing to the upper right, in blue) and the center of the radical Delaunay triangle with the smallest normalized divergence $Q_k$, representing the most significantly underpacked direction (pointing to the upper left, in red).
(c) Probability distribution of $\alpha$. (d) Probability distribution of $\beta$. In (c,d), the curves with the red triangles represent the droplets in less jammed packing with $\phi<0.96$, and the curves with the blue circles represent the droplets in overjammed packings with $\phi \geq 0.96$.  The probability distributions are only for particles with cage-relative displacements $\Delta r > 2.4$~$\mu$m to ensure that the direction is not ill-defined (given that this is our positional uncertainty).
}
\end{figure}

\begin{figure}
\centering
\includegraphics[width=0.8\columnwidth]{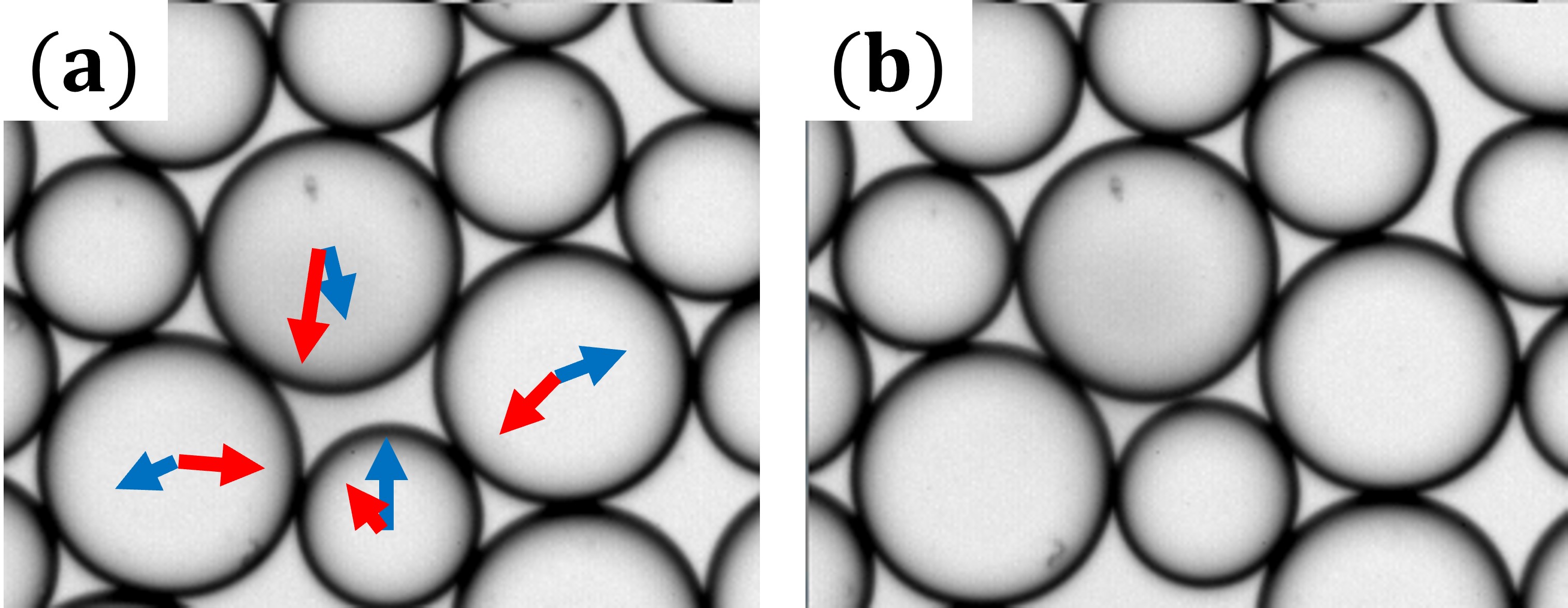}
\caption{\label{t1event} 
One example of a T1-like rearrangement. (a) The particle packing in
experiment.  The red arrows pointing toward the central void demonstrate the Voronoi anisotropy vectors $\vec{C}$. The blue arrows show the displacement of each droplets during the next 30~s:  the top and bottom droplets converge, and the left and right droplets move apart.  Here the converging droplets are already nearest neighbors, so the T1 event (changing nearest neighbors) is more than halfway completed.  (b) The picture of particle packing in experiment with $\delta t = 30s$ after (a).  In this process, two droplets move inward and two move outward.
}
\end{figure}

The discussion thus far has focused on the magnitude of motion; we next turn to considering the direction of motion.  Given that $\vec{C}$ has information about a droplet's neighbors (through the radial Voronoi tessellation), droplets may move in a direction influenced by $\vec{C}$.  As is shown in Fig.~\ref{voronoicell}(b), the anisotropy vector of a droplet $\vec{C}$ typically points towards the free space in its nearest neighborhood. As $\phi$ increases, the droplets tend to fill in the free space and get closer to the neighbors. We calculate the angle $\alpha$ between $\vec{C}$ and the cage-relative displacement of each droplet; see the sketch in Fig.~\ref{direction}(a).  The probability distribution of $\alpha$, as is shown in Fig.~\ref{direction}(c), peaks at ranges of $[0 ^{\circ}, 20 ^{\circ}]$ and $[160 ^{\circ}, 180 ^{\circ}]$ with probability $\approx 0.16$ and $0.14$ respectively.  This is true in both packings that are less jammed (red curve with $\phi<0.96$) and overjammed (blue curve with $\phi \geq 0.96$).  In other words, the droplets are likely to move towards the direction of $\vec{C}$ with $\alpha < 20^{\circ}$, or to the opposite direction with $\alpha > 160^{\circ}$. This experimental phenomenon can be well explained with T1 topological neighbor-exchanging events \cite{weaire84,kraynik88,stavans93,hutzler95,dennin97,kabla03,dennin04,dollet07,weaire10,chen12}. In a T1 rearrangement, two droplets move inward to fill in the free space ($\alpha \approx 0^\circ$), and the other two move outward ($\alpha \approx 180^\circ$), as shown in Fig.~\ref{t1event}.  Thus, not only does a large Voronoi anisotropy vector $\vec{C}$ make it more likely that a droplet rearranges, it also influences the direction of that rearrangement.  The motion with $\alpha \approx 0$ is similar to that observed in prior work which studied motion in a 3D granular packing \cite{slotterback08}; however, they did not observe the $\alpha \approx 180^\circ$ motion, perhaps because T1 events are more complex in 3D materials. In addition to T1 rearrangements, there are numerous motions with $ 20^\circ< \alpha < 160^\circ$. These movements are indicative of nearby droplet motion caused by the T1 rearrangements, which are more random in their direction of motion and generally smaller in magnitude of displacement \cite{desmond15,chen15}.

The $Q_k$'s have directional information like $\vec{C}$ and it is plausible they too should influence the direction of a droplet's displacement.  Among the several Delaunay triangles surrounding a droplet, the one with smallest $Q_{k}$ is the most underpacked region near the droplet. As $\phi$ increases, in general, the droplets move to fill in the underpacked regions. To investigate how $Q_{k}$ relates to the direction of droplet's motion, we define the angle $\beta$ as the angle between $\Delta \vec{r}$ and the direction of the mass center of the radical Delaunay triangle with the smallest $Q_{k}$. This angle is sketched in Fig.~\ref{direction}(b).  The probability distribution of $\beta$, Fig.~\ref{direction}(d), peaks at ranges of $[0 ^{\circ}, 20 ^{\circ}]$ and $[160 ^{\circ}, 180 ^{\circ}]$ with probability $\approx 0.18$ and $0.16$ in both jammed and over-jammed packings. As with the angle $\alpha$ [Fig.~\ref{direction}(c)], this result can also be explained with T1 neighbor-exchanging events.

In terms of directionality, Fig.~\ref{direction} shows that both the first and secondary nearest neighbors influence the direction of droplets movement. Seeing that the peaks of the probability distributions of $\beta$ in Fig.~\ref{direction}(d) are slightly higher than the ones of $\alpha$ in Fig.~\ref{direction}(c), the structure of the secondary nearest neighbors appears to have a more significant impact on the directionality of droplets movement.  Likewise, $\langle 2\cos^2 \alpha - 1 \rangle = 0.16$ and $\langle 2\cos^2 \beta - 1\rangle = 0.26$, a further indication that $\beta$ is more influential.  (This average would be 1 if motion was completely parallel and/or antiparallel to the given direction, and would be 0 if the angle is uniformly distributed.)  We also note that the directions of the Voronoi anisotropy vector and the direction of the smallest $Q_k$ are similar:  the distribution of the difference between their angles has a mean of $0^\circ$ and a standard deviation of $48^\circ$.

\section{Conclusion}

Our goal has been to understand how local structure influences rearrangements in a quasi-two-dimensional emulsion where the area fraction slowly increases toward $\phi = 0.99$.  To do this, we use the Voronoi polygon analysis introduced by Rieser {\it et al.} \cite{rieser16}.  In this article they present one measure that has information about a particle and its nearest neighbors; and a secondary measure that includes information from the second nearest neighbors as well.  These two measures characterize spatial heterogeneity, for example the presence of larger voids near a particle, or the presence of a more densely packed region nearby.  We find both of these measures are correlated with particle motion:  larger spatial heterogeneity predicts subsequent larger particle displacements.  Moreover, both measures are vectors pointing toward voids.  These directions preferentially orient the T1 rearrangements, giving a sense of where droplets can converge together (to fill in the void) or move apart (thus making room for the converging droplets).  These two measures of spatial heterogeneity are correlated, so it makes sense that they are both effective predictors of motion.  That being said, the measure accounting for first nearest neighbors is slightly more predictive of a particle's likelihood of rearranging, while the measure accounting for first and second nearest neighbors is slightly more predictive of the particle's direction of motion.

In our experiment the sample changes area fraction from just mildly jammed ($\phi = 0.88$) to nearly confluent ($\phi = 0.99$).  It is intriguing that the relationship between the structural measures and the particle displacements remains consistent over the entire area fraction range.  During the experiment the mean value of one measure drops by more than factor of 2 [Fig.~\ref{fig:phivstime}(c)] and the displacements drop by more than a factor of 10 [Fig.~\ref{fig:phivstime}(b)], yet the correlations between structure and displacement magnitude are qualitatively the same [Fig.~\ref{drcage}(a,b)], and correlations between structure and displacement direction are quantitatively the same [Fig.~\ref{direction}(c,d)].  It is plausible that machine learning techniques could discern structural measures with even stronger predictive power \cite{cubuk17,bapst20,boattini21}.  Our results suggest that for machine learning algorithms, it would be useful to use area-fraction independent structural quantities for the machine learning analysis.  Our results also suggest that key structural features of importance are the spatial heterogeneity of $\phi$ and in particular the voids.

It is also noteworthy that the center of the radical Voronoi polygon matters even as $\phi \rightarrow 1$ and the voids vanish to points.  At these high area fractions it is plausible that droplets minimize their surface energy when their center of mass lies closest to the center of mass of their radical Voronoi polygon; this generally leads to more compact polygons \cite{klatt19}.  Thus it is sensible that droplet rearrangements will still be influenced by the structural quantities we consider.  This bears similarities to model systems such as ``Voronoi fluids'' \cite{ruscher16} and the ``geometric Lloyd's algorithm'' \cite{lloyd82,klatt19,hain20} which study collections of generating points and describe the energy of the system in terms of the Voronoi polygon shapes made from these points.  Other similar models are used to study the rearrangements of confluent cells at $\phi=1$, for which a variety of dynamics are also possible \cite{bi14,bi16,sussman17cellgpu}.  In the case of Lloyd's algorithm, the goal is to move the generating points such that they move toward the center of mass of their Voronoi polygon \cite{klatt19}, thus matching our results by construction.  For the other dynamics, one can study equilibrated systems for which rearrangements do not change the distributions of $\vec{C}$ or $\sigma_{\rm div}$, the structural quantities we study.  That being said, it is still plausible that these structural quantities may influence the larger displacement rearrangements in these $\phi=1$ Voronoi-based models.

Acknowledgements:  We thank X.~Hong, C.~Orellana, J.~Rieser, and D.~Sussman for helpful discussions.  This material is based upon work supported by the National Science Foundation under Grant No. CMMI-1250199/-1250235 (X.D.) and CBET-2002815 (E.R.W.).

\bibliography{eric}
\end{document}